# 36 Hz integral linewidth laser based on a photonic integrated 4.0-meter coil resonator

KAIKAI LIU,[1,7] NITESH CHAUHAN,[1,7] JIAWEI WANG,[1] ANDREI ISICHENKO,[1] GRANT M. BRODNIK,[1] PAUL A. MORTON,[2] RYAN BEHUNIN,[3,4] SCOTT B. PAPP,[5,6] AND DANIEL J. BLUMENTHAL[1,*]

[1]Department of Electrical and Computer Engineering, University of California Santa Barbara, Santa Barbara, CA, USA
[2]Morton Photonics, West Friendship, MD, USA
[3]Department of Applied Physics and Materials Science, Northern Arizona University, Flagstaff, Arizona, USA
[4]Center for Materials Interfaces in Research and Applications (¡MIRA!), Northern Arizona University, Flagstaff, AZ, USA
[5]Department of Physics, University of Colorado Boulder, Boulder, CO, USA
[6]Time and Frequency Division 688, National Institute of Standards and Technology, Boulder, CO, USA
[7]Equal contribution authors
*danb@ucsb.edu

**Abstract:** Laser stabilization sits at the heart of many precision scientific experiments and applications, including quantum information science, metrology and atomic timekeeping. These systems narrow the laser linewidth and stabilize the carrier by use of Pound-Drever-Hall (PDH) locking to a table-scale, ultra-high quality factor (Q), vacuum spaced Fabry-Perot reference cavity. Integrating these cavities, to bring characteristics of PDH stabilization to the chip-scale, is critical to reduce their size, cost, and weight, and enable a wide range of portable and system-on-chip applications. We report a significant advance in integrated laser linewidth narrowing, stabilization and noise reduction, by use of a photonic integrated 4.0-meter-long coil resonator to stabilize a semiconductor laser. We achieve a 36 Hz $1/\pi$-integral linewidth, an Allan deviation (ADEV) of $1.8\times10^{-13}$ at 10 ms measurement time, and a 2.3 kHz/sec drift, to the best of our knowledge the lowest integral linewidth and highest stability demonstrated for an integrated reference cavity. Two coil designs, stabilizing lasers operating at 1550 nm and 1319 nm are demonstrated. The resonator is bus coupled to a 4.0-meter-long coil, with a 49 MHz free spectral range (FSR), a mode volume of $1.0\times10^{10}$ μm$^3$ and a 142 million intrinsic Q, fabricated in a CMOS compatible, ultra-low loss silicon nitride waveguide platform. Our measurements and simulations show that the thermorefractive noise floor for this particular cavity is reached for frequencies down to 20 Hz in an ambient environment with simple passive vibration isolation and without vacuum or thermal isolation. The TRN limited performance is estimated to be an 8 Hz $1/\pi$ integral linewidth and ADEV of $5\times10^{-14}$ at 10 ms, opening a stability regime that heretofore has only been available in fundamentally un-integrated systems. Our results demonstrate the potential to bring the characteristics of lab-scale stabilized lasers to the integrated, wafer-scale compatible chip-scale, enabling portable and lower cost atomic clocks and navigation, microwave photonics, quantum applications, and energy-efficient coherent communications systems.

## 1. INTRODUCTION

Spectral purity is critical for applications that demand low laser phase noise and high carrier stability, including atomic clocks [1], microwave photonics [2,3], quantum applications [4,5], and energy-efficient coherent communications systems [6]. Lasers with integral linewidth below 1 Hz and an Allan deviation better than $10^{-16}$ over ~1 second are realized by use of Pound-Drever-Hall (PDH) locking to ultra-stable optical cavities [7–11]. This level of performance is achieved with table-top vacuum-cavity, ultra-high finesse resonators that employ sophisticated mirror and athermal designs, and other features to mitigate intrinsic thermal fluctuations [12,13]. Translating qualities of these systems to a photonic integrated, CMOS compatible wafer-scale platform will enable new generations of precision portable applications as well as systems-on-chip integration [14–16].

Progress towards miniaturization of PDH-locked stabilized lasers has centered mainly on bulk-optic reference cavities and a limited number of integrated cavities. Centimeter-scale whispering gallery mode resonators have yielded a fractional frequency noise (FFN) stability of $6\times10^{-14}$ at 100 ms in a thermally and acoustically isolated vacuum enclosure [17]. A fused silica micro-cavity stabilized a semiconductor laser to a 25 Hz integral linewidth and a FFN stability of $1\times10^{-13}$ at 20 ms [18,19]. In general it is desirable to increase the optical mode volume to reduce the various intrinsic thermal noise that scales inversely with mode volume [17,20–22]. Stabilization of a fiber laser to a deep-etched silica waveguide spiral resonator yielded $4\times10^{-13}$ at 0.4 ms [21] and on the order of 100 Hz linewidth; however, new approaches that are less environmentally sensitive and wafer-scale CMOS compatible are critical. Recent CMOS compatible, silicon nitride resonators achieved a $1/\pi$-linewidth reduction from 3.66 kHz to 292 Hz and a carrier stability of $6.5\times10^{-13}$ at 8 ms [23].

In this paper, we report a significant advance in photonic integrated laser stabilization, achieved by locking semiconductor lasers, in both C- and O-band, to 4.0-meter-long photonic integrated coil resonators. We measure a 36 Hz $1/\pi$-integral linewidth, a 345 Hz β-separation linewidth [24], an Allan deviation of $1.8\times10^{-13}$ at 10 ms and a 2.3 kHz/sec drift, the lowest integral linewidth and highest stability reported for an integrated reference cavity, to the best of our knowledge. Both the 1550 nm and 1319 nm coil resonators employ a 4.0-meter-long round-trip, significantly lowering the cavity-intrinsic thermorefractive noise over other designs [22] as well as susceptibility to photothermal (PT) noise induced fluctuations. The 1550 nm coil resonator measures a 49.1 MHz FSR, an intrinsic 80 Million Q, a loaded 55 Million Q and has a ~$1.0\times10^{10}$ μm$^3$ mode volume and the 1319 nm coil resonator a 48.9 MHz FSR, an intrinsic 142 Million Q, a loaded 71 Million Q and a ~$0.6\times10^{10}$ μm$^3$ mode volume. We also present numerical modeling of the PDH locking loop noise and the PT and TRN contributions. We demonstrate that the stabilized frequency noise reaches the thermorefractive noise (TRN) floor down to 20 Hz frequency offset from carrier. The TRN limited performance is estimated to be an 8 Hz $1/\pi$ integral linewidth and ADEV of $5\times10^{-14}$ at 10 ms, indicating that with mitigation of other noise sources, further performance improvement is possible. We compare the frequency noise and drift to a Brillouin laser locked to a fused silica microrod Fabry-Perot reference cavity with 25 Hz integral linewidth and show that our linewidth, frequency noise and ADEV performance are comparable. Below 20 Hz, environmental noise dominates, as the resonator is located in an ambient environment using simple passive vibration damping isolation and without vacuum or thermal isolation. These results open the door to integrated all-waveguide frequency stabilized lasers for portable atomic clocks, microwave photonic, quantum applications, precision spectroscopy, and energy-efficient coherent communications systems.

## 2. RESULTS

The $Si_3N_4$ coil waveguide resonator employs a high-aspect ratio waveguide core design, 6 μm wide by 80 nm thick, with a 15 µm thick thermally silicon dioxide lower cladding is and 6 µm thick oxide upper cladding deposited by tetraethoxysilane pre-cursor plasma-enhanced chemical vapor (TEOS-PECVD). Such a waveguide dimension is chosen with the considerations of using both the high-aspect ratio waveguide and the fundamental transverse magnetic (TM) mode to mitigate scattering loss, minimizing the coil bending radius and waveguide bending loss, and minimizing cross-talk between coil waveguides [25–27]. The coil waveguide spacing is 40 µm and the minimum bending radius is 9.0 mm. The only difference in resonator design between the 1550 nm and 1319 nm devices is the bus-to-resonator coupler, which for 1550 nm is a directional coupler with a 3.5 μm gap and 3.0 mm coupling length and for 1319 nm is a directional coupler with a 3.5 µm gap and 0.5 mm coupling length. Such a coupler design gives proper coupling for the fundamental TM mode and makes weak coupling for the transverse electric modes [28]. The design of the coil waveguide and the directional coupler is described in further detail in Supplement 1 Section 1.

To measure the resonator Qs, we use the fiber-extended-cavity semiconductor laser developed by Morton Photonics [2] to probe the resonator and carry out spectral scan with a radio frequency (RF) calibrated fiber Mach-Zehnder interferometer (MZI) as an optical frequency ruler. The 1550 nm resonator is measured to have an intrinsic 80 Million and 55 Million loaded Q (Fig. 1(b)), corresponding to a propagation loss of 0.39 dB/m; the 1319 nm resonator is measured to have an intrinsic 142 Million and 71 Million loaded Q (Fig. 1(c)), corresponding to a propagation loss of 0.16 dB/m.

The laser stabilization setup using the Morton Photonics laser and frequency noise measurement methods are shown in Fig. 2, where the resonators are mounted on active temperature controlled stages, inside passive enclosures on a floating optical table, the same as described in our previous work [23]. With the measured resonator Qs, the PDH lock only needs ~0.2 mW optical power onto the photodetector to gain enough signal-to-noise ratio (SNR) against various noise sources and the PDH locking bandwidth is about 1 MHz (see Supplement 1 Section 2 for the detailed PDH lock setup). To measure the frequency noise and laser carrier stability, we use two independent methods. A fiber unbalanced Mach-Zehnder interferometer (MZI) with a 1.026 MHz FSR is used as an optical frequency discriminator (OFD) for self-delayed homodyne laser frequency noise measurement at frequency offset greater above 1 kHz [23,29,30], as the fiber noise in the MZI dominates at lower frequencies; for frequency noise below 1 kHz offset, we employ a Rock™ single frequency fiber laser that is PDH locked to a Stable Laser Systems™ (SLS) ultra-low expansion (ULE) cavity and produces a Hz-level linewidth output at 1550 nm wavelength with a frequency drift of ~0.1 Hz/s, referred to as the stable reference laser (SRL), as a frequency noise reference below 1 kHz offset. The SRL frequency noise measurement below 1 kHz is done by photomixing the SRL with our stabilized laser on a high-speed photodetector to produce a heterodyne beatnote signal and recording the beatnote frequency on a frequency counter. The OFD and SRL frequency noise measurements are discussed in detail in Supplement 1 Section 3.

The frequency noise measurements for the 1550 nm coil-resonator stabilized laser are shown in Fig. 3, where the OFD and SRL frequency noise measurements are stitched at 1 kHz frequency offset to make a complete frequency noise measurement for both the free-running and stabilized laser. From the frequency noise spectrum, $1/\pi$-integral linewidth and $\beta$-separation linewidth can be calculated to evaluate the laser noise and linewidth performance [11,24]. The $1/\pi$-integral linewidths for the free-running and stabilized laser in Fig. 3(a) are 1.3 kHz and 36 Hz, respectively; Besides, the $\beta$-separation linewidths for the free-running and stabilized laser are also calculated to be 91 kHz and 345 Hz, respectively (see Supplement 1 Section 3 and Fig. S4 for $1/\pi$-integral and the $\beta$-separation linewidth calculations). The Allan deviation calculated from the OFD and SRL frequency noise measurements for the stabilized laser reaches a minimum of $1.8 \times 10^{-13}$ at 10 ms. The stabilized laser's frequency noise spectrum falls onto the resonator-intrinsic thermorefractive noise from 50 kHz down to 20 Hz offset, as shown in Fig. 3(a), whereas below 20 Hz the resonator in the ambient environment could suffer from environmental noise such as thermal fluctuation noise. We also demonstrate this coil resonator laser stabilization at O band 1319 nm as shown in Fig. 4, where the stabilized laser measured by OFD reaches the thermorefractive noise limit from 1 kHz to 50 kHz. Fig. 3(c) shows the direct measurement of the coil resonator's linear drift of 2.3 kHz/s, a 100X reduction compared to the free-running laser's 193 kHz/s. In Fig. 3(a), we also show the frequency noise of a stimulated Brillouin scattering (SBS) laser [30] locked to a Billion Q fused microrod Fabry-Perot cavity [19] and a fiber laser locked to the SLS cavity to show that our coil resonator stabilized laser achieves comparable performance.

To understand and analyze the 1550 nm stabilized laser frequency noise, we simulate and model various noise sources in the PDH lock such as photodetector noise and shot noise in the locking loop and photothermal noise and thermorefractive noise in the coil resonator. Qs of reference cavities for laser stabilization are a critical parameter in determining the stabilized laser performance, as the SNR for the frequency discrimination against noise sources such as photodetector noise, shot noise and electronics noise is proportional to Q and optical power onto the photodetector [31]. With the measured resonator Qs, ~0.2 mW optical power onto the photodetector and the noise figures of the photodetector used for this purpose (Thorlabs PDB470C), we can estimate that the photodetector noise of $4 \times 10^{-3}$ Hz$^2$/Hz and the shot noise of $6 \times 10^{-4}$ Hz$^2$/Hz (see Supplement 1 Section 2 for the frequency noise calculation), which do not pose a limit on the PDH lock performance. As both photothermal noise and thermorefractive noise scales with 1/(optical mode volume) or 1/(round-trip length) and the coil resonator has a 4.0 meter long round-trip length, the photothermal noise is estimated not to be a dominant contribution (see Supplement 1 Section 2 for calculation details) and the thermorefractive noise modelled using the method in Ref. [32,33] reaches as low as 1.2 Hz$^2$/Hz at 1 Hz offset (green dash curve in Fig. 3(a)). Therefore, all these noise sources are below the thermorefractive noise below 1 MHz frequency offset. However, below 20 Hz offset it is very likely that the environmental noise becomes dominant. Using techniques to reduce the environmental noise of the coil resonator, the stabilized laser frequency noise spectrum can tend towards a thermorefractive-noise-limited integral linewidth of 8 Hz.

## 3. DISCUSSION

We report a significant advance in photonic integrated laser stabilization, achieved by locking a fiber-extended cavity semiconductor laser to a 4.0-meter-long round-trip coil waveguide resonator with a free spectral range (FSR) of ~49 MHz. We demonstrate that the stabilized laser frequency noise reaches the resonator thermorefractive noise limit from 50 kHz down to 20 Hz and we measure a $1/\pi$-integral linewidth of 36 Hz and β-separation linewidth of 345 Hz and Allan deviation of $1.8\times10^{-13}$ at 10 ms using the 1550 nm coil resonator, which we believe is the lowest linewidth and highest stability reported for an all-waveguide design, to the best of our knowledge. Reaching the thermorefractive noise limit is also demonstrated at O band 1319 nm. Further improvements in laser spectral purity can be achieved with coil length optimization, resonator packaging for environmental noise isolation. Recently, we reported an integrated $Si_3N_4$ transverse electric (TE) mode waveguide resonator with an intrinsic Q of 422 Million [27] and an integrated $Si_3N_4$ transverse magnetic (TM) mode waveguide resonator with an intrinsic Q of 720 Million at C band [28]. Increasing the resonator Qs gives us higher SNR for the PDH lock. The results presented in this paper demonstrate the potential to bring the performance of table-top and miniaturized ultra-high Q resonators to planar all-waveguide solutions and benefit compact versions of applications that require laser spectral purity such as atomic clocks, microwave photonics, quantum applications, and energy-efficient coherent communications systems.

These results are an important milestone for integrated stabilization, representing the longest integrated coil resonator to date, that approaches the performance of bulk-optic fused silica microcavity Fabry-Perot resonator stabilizing a semiconductor laser to 25 Hz linewidth and $1\times10^{-13}$ ADEV at 20 ms [18,19]. Normalizing the $1.8\times10^{-13}$ ADEV to the 193 THz carrier yields a 36 Hz linewidth, consistent with the integrated 36 Hz integral linewidth. The internal consistency of these distinct measurements enables us to determine that the carrier jitter is the major contribution to the 36 Hz integral linewidth. Additional linewidth broadening due to lower frequency terms and drift are also considered; we measure a 345 Hz β-separation linewidth and 2.3 kHz/second drift rate, which are also record low numbers for an integrated resonator technology.

The TRN simulation for this resonator design is based on assumed optical mode parameters and the material thermal parameters, and as such only gives a range of estimates for the coil resonator TRN and FN spectrum. The actual values are sensitive to and highly dependent on fabrication and material parameters including conductivity, density, heat capacity, refractive index, and thermo-optic coefficient. As such, our analysis is a best estimate based on available information, and future work includes working to understand and identify these parameters and their contributions in more detail. Additionally, the TRN doesn't explain other features like the flat FN spectrum from 10 kHz to 100 kHz and future work will involve understanding the noise mechanisms that contribute to this frequency band. Based on our estimates, the TRN limited performance is an 8 Hz $1/\pi$ integral linewidth and ADEV of $5\times10^{-14}$ at 10 ms, indicating that with mitigation of other noise sources, further performance improvement is possible.

Further improvements can be achieved by increasing the resonator Q [27,28] and improving the resonator packaging. Increased Q will result in a higher SNR for the PDH lock and improved packaging can reduce the environmentally induced frequency noise. Other areas of improvement include increase in coil length and incorporation of athermal designs [34] and incorporating dual waveguide mode precision temperature measurements [35] and these cavities can be combined in parallel or serial for feedback loop noise engineering. These results show promise to bring the characteristics of table-top and miniaturized ultra-high Q resonators to planar all-waveguide solutions, that are CMOS wafer-scale compatible, and bring benefits to applications for a wide range (including mobile) precision application including atomic clocks, microwave photonics, quantum sensing, metrology and computing applications, and energy-efficient coherent communications systems.

**Funding.** This material is based upon work supported by the Advanced Research Projects Agency-Energy (ARPA-E), U.S. Department of Energy, under Award Number DE-AR0001042, and NIST. The views and conclusions contained in this document are those of the authors and should not be interpreted as representing official policies of DARPA, ARPA-E or the U.S. Government.

**Disclosures.** The authors declare no conflicts of interest.

**Data availability.** Data underlying the results presented in this paper are not publicly available at this time but may be obtained from the authors upon reasonable request.

**Supplemental documents.** See Supplement 1 for supporting content.

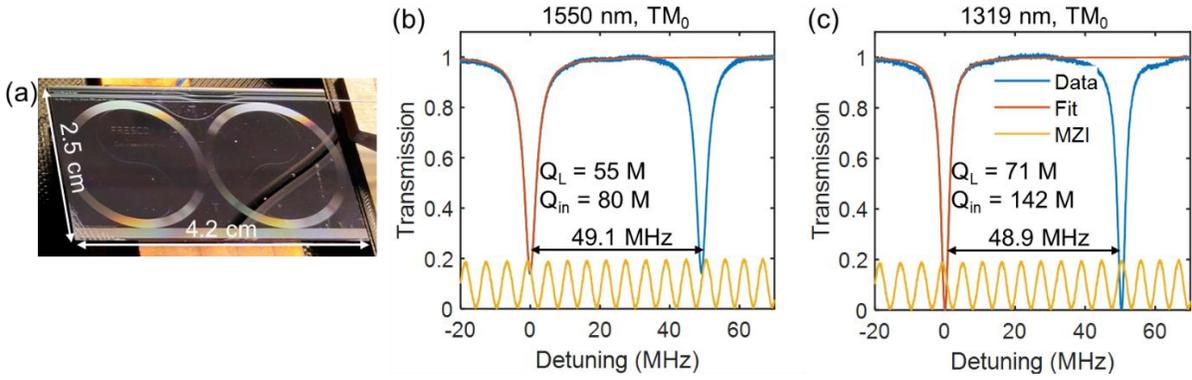

**Fig. 1.** Coil waveguide resonator design and Q. (a) Actual device picture and its dimension. (b) C-band (1550 nm) coil resonator with an FSR of 49.1 MHz and an intrinsic 80 million and loaded 55 million Q. (c) O-band (1319 nm) coil resonator with an FSR of 48.9 MHz and an intrinsic 142 million and loaded 71 million Q.

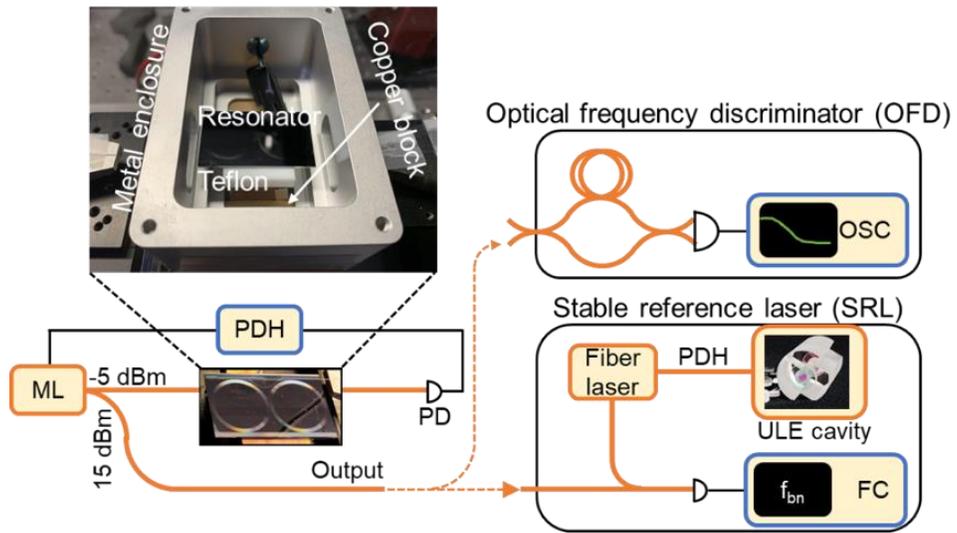

**Fig. 2.** PDH lock setup and frequency noise measurements for a Morton Photonics laser (ML) locked to a coil resonator in an ambient environment, where the resonator sits on a Teflon block and a thermal-electric cooler underneath the copper block provides an mK-level temperature stability. The OFD measures the laser's frequency noise above 1kHz offset frequency, and a heterodyne signal from the photomixing of the laser and the SLS laser is sent to the frequency counter (FC) for frequency noise measurement below 1kHz offset frequency. OSC, oscilloscope; PD, photodetector.

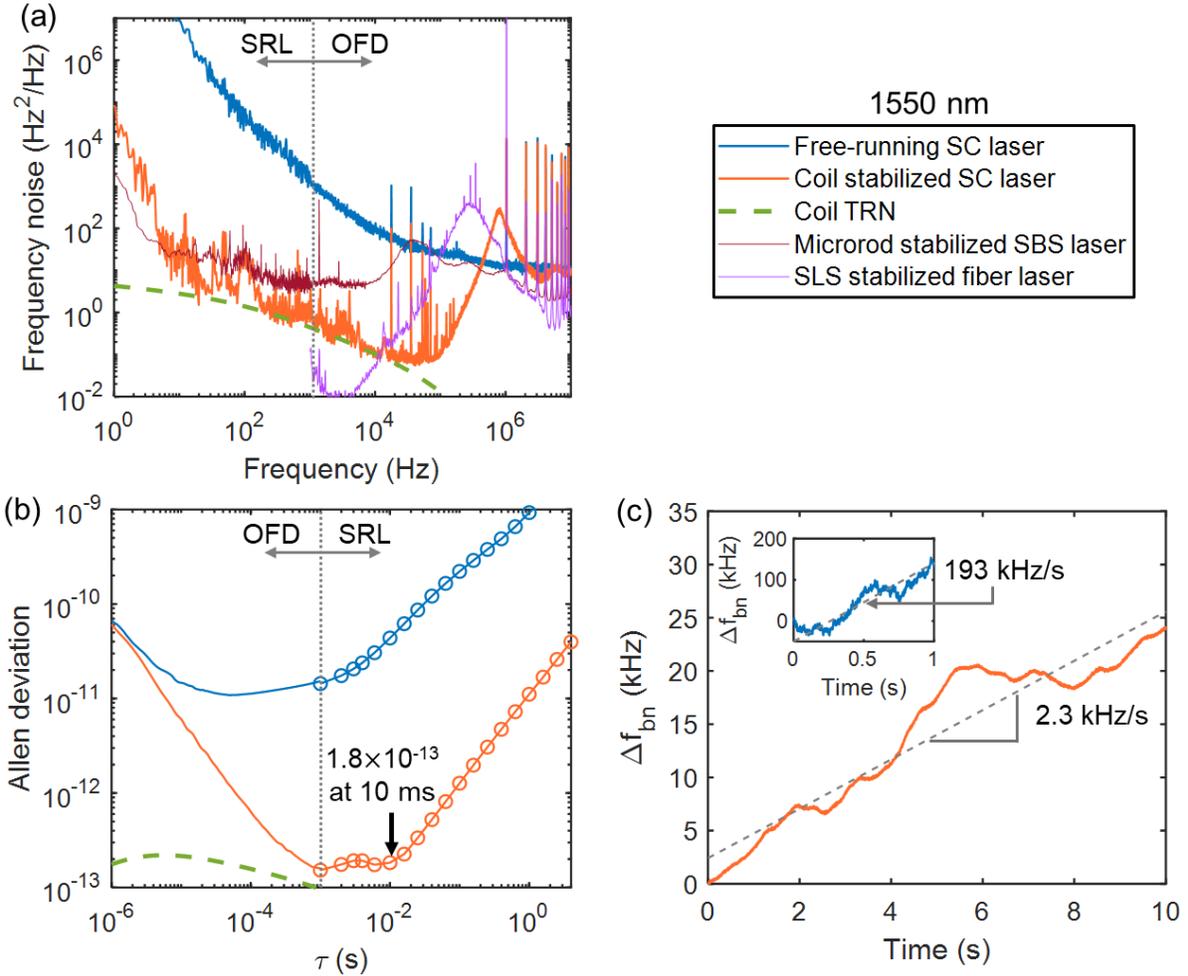

**Fig. 3.** 1550 nm integrated coil resonator stabilized laser frequency noise, Allan deviation and beatnote frequency drift. (a) OFD and SRL frequency noise measurements for the free-running and coil stabilized laser and comparison to performance of an SBS laser stabilized to a bulk-optic fused silica microrod resonator. SRL frequency noise above 1 kHz is included to show properties of laser used for heterodyne beatnote measurements indicated by SRL region of noise plot. OFD regions denote frequency noise data taken with a fiber optic Mach-Zehnder interferometer frequency discriminator and SRL regions denote frequency noise data taken by beating the coil-stabilized laser with a ~ 1 Hz linewidth cavity stabilized reference laser. (b) Allan deviation is estimated based on the heterodyne signal's frequency recorded by a frequency counter at an averaging time τ above 1 ms; Allan deviation is calculated from the measured OFD frequency noise at averaging time τ below 1 ms. The dashed light orange curve is calculated with the linear drift removed. The stabilized laser measures an Allan deviation of $1.8×10^{-13}$ at 10 ms. (c) Beatnote frequency drift ($\Delta f_{bt} = f_{bn}(t) - f_{bn}(0)$) is measured on the frequency counter with a gatetime of 1 ms, where $f_{bn}(0)$ is ~54 MHz for the free-running and ~61 MHz for the stabilized laser. OFD, optical frequency discriminator; SRL, stable reference laser; SC laser, semiconductor laser; TRN, thermorefractive noise; SBS, Stimulated Brillouin Scattering.

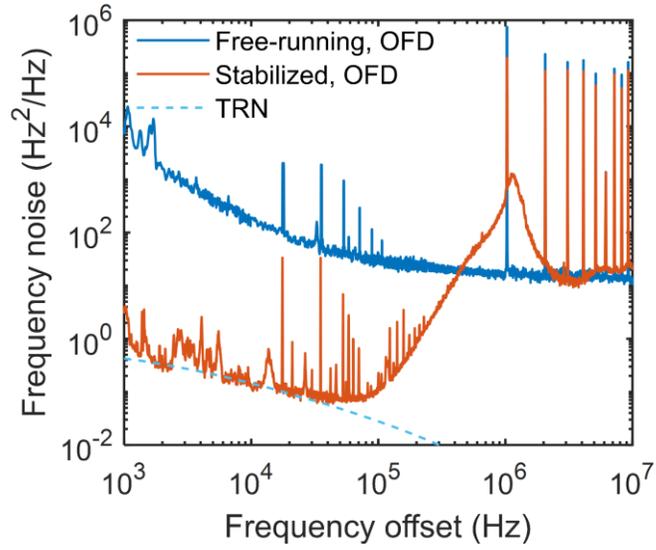

**Fig. 4.** O-band 1319 nm laser stabilization with the coil waveguide resonator and frequency noise measured by the optical frequency discriminator (OFD) method. The stabilized laser reaches the thermorefractive noise limit from 1 kHz to 50 kHz.

# 36 Hz integral linewidth laser based on a photonic integrated 4.0-meter coil resonator: Supplement

## 1. COIL WAVEGUIDE RESONATOR DESIGN

The waveguide dimension is 6 μm by 80 nm. The bus-to-ring coupling is a directional coupler with a gap of 3.5 μm and length of 0.5 mm for 1550 nm (3 mm for 1319 nm). The fabrication processes of these resonators include deposition of 40 nm thick $Si_3N_4$ thin film using low-pressure chemical vapor deposition (LPCVD) on 15 μm thermal oxide grown on the silicon wafer, the etching process, the 6 μm upper oxide cladding deposition using tetraethoxysilane pre-cursor plasma-enhanced chemical vapor deposition (TEOS-PECVD), and the final 11 hours' annealing in ~1100 C, which is described in detail in our previous publication [1].

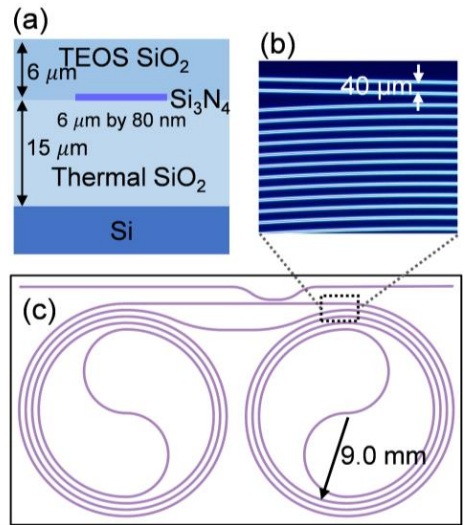

**Fig. S1.** Coil resonator design. (a) The resonator waveguide is a high-aspect-ratio design, 6 μm wide by 80 nm thick. (b) Microscope image of the coil waveguides. (c) Layout diagram of the coil resonator. The minimum bending radius is 9.0 mm and the spacing between the waveguides is 40 μm.

## 2. PDH LOCK NOISE ANALYSIS AND FREQUENCY NOISE MEASUREMENTS

The frequency discrimination slope using a critically coupled resonator with a certain linewidth $\delta\nu$ and Q factor is expressed as [2],

$$\epsilon = \frac{8\sqrt{P_c P_s}}{\delta\nu}, \tag{S1}$$

where $P_c$ is the optical carrier power and $P_s$ the optical sideband power. Noise sources such as the photodetector noise and shot noise usually pose limits on the PDH lock performance. The shot noise equivalent frequency noise can be expressed as,

$$S_{sh} = \frac{\delta\nu}{4}\sqrt{\frac{h\nu}{P_c}}, \tag{S2}$$

and photodetector noise equivalent frequency noise can be expressed as,

$$S_{sh} = S_{NEP}\frac{\delta\nu}{8P_c P_s}, \tag{S3}$$

where $S_{NEP}$ is the noise equivalent power (NEP) of the photodetector used in the PDH lock. In our PDH lock experiments, the sideband power is ~0.1 of the carrier power with a total optical power of ~0.2 mW and the Thorlabs PDB470C photodetector with a conversion gain of 10 kV/W and an

NEP of 8 pW/$\sqrt{Hz}$ is used. With 0.2 mW optical power onto the photodetector and 55 Million cavity Q, the photodetector noise equivalent frequency noise is calculated to be $4\times10^{-3}$ Hz$^2$/Hz and the shot noise equivalent frequency noise is calculated to be $6\times10^{-4}$ Hz$^2$/Hz.

Using the thermo-optic coefficients of SiO2 ($0.95\times10^{-5}$ 1/K) and SiN ($2.45\times10^{-5}$ 1/K) at 1550 nm reported in the literature [3,4], we perform a Comsol simulation that simulates the thermal heating due to absorption heating and estimates the redshift from optical absorption heating: $\alpha_0 = \Delta f_{res}/P_{abs}$ =0.11 MHz/mW. Since the optical power onto the photodetector is ~0.2 mW and the fiber-to-chip insertion loss is ~3 dB, the on-chip optical power in the coil resonator is $P$ =~0.4 mW. Assuming that the laser RIN is flat and -120 dBc/Hz and $\xi$ = 10% of the on-chip optical power contributes to absorption heating, the resulting photothermal noise can be estimated using $S_{PT} = RIN(\xi P\alpha_0)^2$, which turns out to be $2\times10^{-5}$ Hz$^2$/Hz. More rigorously speaking, the resonance frequency shifting response from absorption heating has a roll-off above ~1 kHz and can be expressed as $\alpha(f) = \alpha_0 H(f)$, where $H(f)$ characterizes the response at different frequencies with $H(0) = 1$. Therefore, the photothermal noise can be expressed as $S_{PT}(f) = H(f)^2 \times 2\times10^{-5}$ Hz$^2$/Hz with $S_{PT}(0)$ =$2\times10^{-5}$ Hz$^2$/Hz as an upper bound for this estimate.

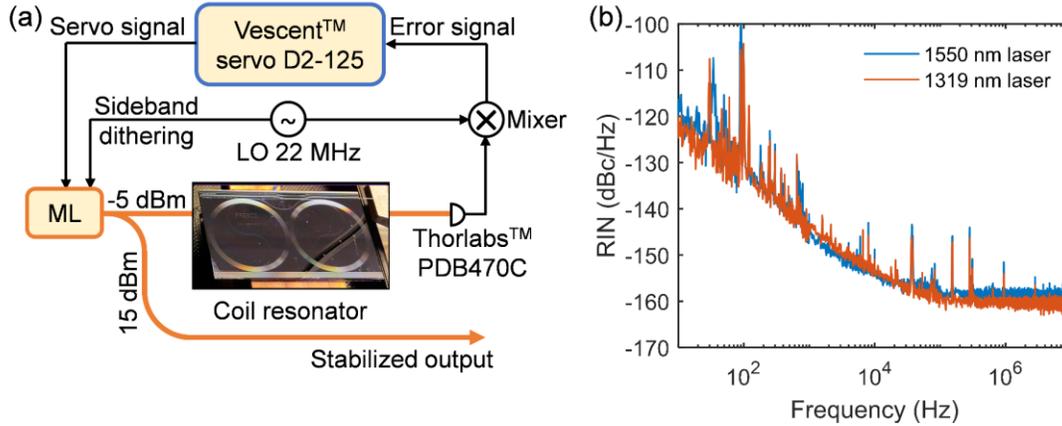

**Fig. S2.** PDH lock set and laser RIN. (a) Detailed setup of the PDH lock. ML, Morton laser developed by Morton Photonics. (b) Relative intensity noise measured by OEWaves™ laser noise measurement instrument.

## 3. FREQUENCY NOISE MEASUREMENT METHODS AND LASER LINEWIDTH

A single-frequency Rock™ fiber laser PDH-locked to and stabilized by a commercially available Stable Laser System™ ultra-low expansion cavity with a ~400,000 finesse, a several-kHz cavity linewidth and a stability below $10^{-15}$ at 1 s averaging time scale, which is referred to as the stable reference laser (SRL), is used for laser frequency noise measurement below 1 kHz frequency offset. The beatnote between the stabilized laser and the SRL is detected by a high speed Thorlabs DET01CFC photodetector. The beatnote frequency is around 5~100 MHz. We employ the Keysight 53200A frequency counter to read the beatnote frequency to record the beatnote frequency time traces with several different gate times (0.01 ms, 0.1 ms and 1 ms) or sampling rates. Frequency noise spectrum is calculated by estimating the single-sided power spectral density of the time traces of the beatnote frequency. Such a frequency noise measurement using the SRL is referred to as the SRL frequency noise measurement for below 1 kHz offset. Above 1 kHz offset, we use an unbalanced fiber Mach-Zehnder interferometer (MZI) with a FSR of 1.027 MHz as an optical frequency discriminator (OFD) for self-delayed homodyne frequency noise measurement, which we refer to as the OFD frequency noise measurement [1,5,6]. To show that the SRL is quite enough to make the laser frequency noise below 1 kHz and the MZI's excessive frequency noise at below 1 kHz frequencies the OFD measurement, the OFD measurements of both the coil resonator stabilized 1550 nm laser and the SRL are shown in Fig. S3. The SRL reaches below $1\times10^{-2}$ Hz$^2$/Hz at 3 kHz offset and below the coil resonator TRN limit and, therefore, is quite enough for noise measurement below 1 kHz offset. Yet, the increasing frequency in the SRL OFD frequency noise at 1 kHz and below is contributed from the OFD MZI, resolving the MZI's excessive noise at low frequencies. In Fig. S3, the OFD frequency noise spectra match with each other below 1 kHz, revealing the same OFD MZI excessive noise.

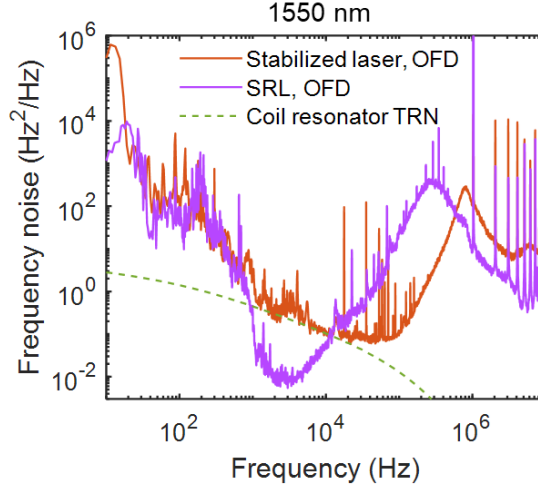

**Fig. S3.** OFD frequency noise measurements on the coil resonator stabilized laser and the SRL show that the SRL is quite enough below 3 kHz offset and the excessive noise from MZI in the OFD measurements.

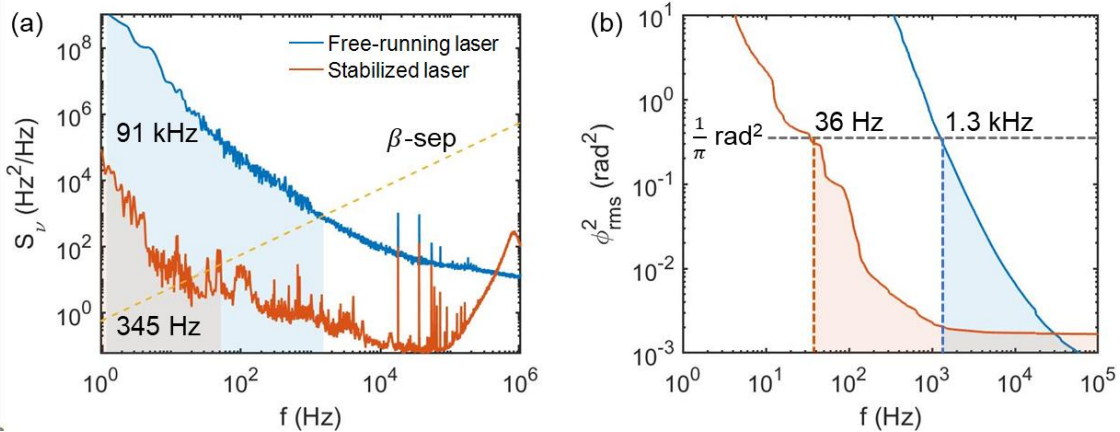

**Fig. S4.** 1/π-integral and β-separation linewidth calculations for the free-running and stabilized laser.

The 1/π-integral linewidth $\Delta\nu_{1/\pi}$ calculated by integrating the single sided phase noise from the highest measured frequency offset (10MHz) to the frequency offset at which the integral is 1/π rad$^2$ is a metric that measures the low-frequency noise [7],

$$\phi_{rms}^2 = \int_{\Delta\nu}^{\infty} S_\varphi(\nu)df = \frac{1}{\pi}. \tag{S4}$$

The β-separation linewidth $\Delta\nu_\beta$ is calculated by integrating the frequency noise spectrum from the lowest frequency offset $1/T_0$ that is above the β-separation line [7],

$$\Delta\nu_\beta = \int_{1/T_0}^{\infty} H(S_\nu(f) - 8\ln(2)f/\pi^2)S_\nu(f)df. \tag{S5}$$